\documentclass[aps,prl,reprint,groupedaddress]{revtex4-1}
\usepackage{graphics}
\usepackage{graphicx}
\usepackage{epsfig}
\begin{document}

\title{Pairing gaps and Fermi energies at scission for $^{296}$Lv alpha-decay }

\author{M. Mirea}

\affiliation{Horia Hulubei National Institute for Physics and Nuclear Engineering,
P.O. Box MG-G, Bucharest, Romania}

%Pairing density-dependent delta interaction  within the two center shell model

\begin{abstract}
The pairing corrections, the single particle occupation numbers, are
investigated within density-dependent delta interaction formalism for
pairing residual interactions. 
The potential barrier is computed
in the framework of the macroscopic-microscopic model.
The microscopic part is based on the Woods-Saxon two center shell model.
The $\alpha$-decay of a superheavy
element is treated, by paying a special attention to the region
of the scission configurations.
The sequence of nuclear shapes follows the superasymmetric fission path for
alpha decay. It was found that the pairing gaps of the states that reach asymptotically the
potential well of the alpha particle have large values at scission 
but become zero after scission.
The 1s$_{1/2}$ single particle levels of the nascent $\alpha$ particle
are fully occupied while the superior levels are empties in the
scission region and remains in the same states during the penetration
of the Coulomb barrier. The projection of the numbers of particle on the
two fragments are obtained naturally. At scission, the nascent $\alpha$ particle forms
a very bound cluster.
\end{abstract}

\maketitle

\section{Introduction}

As pointed out in Ref. \cite{younes}, the disentanglement of the wave
functions at scission between two independent fragments
is an essential ingredient for calculations of the dynamical
observables that characterize the fission. 
The Pauli principle involves
a correlation between the fission fragments even after their separation.
The condition of conservation for the
number of particles in the BCS theory implies the
existence of only one value of  the Fermi energy in the
precise moment when the nucleus breaks. 
The total number of nucleons of the system should be equal with the sum
of the occupations probabilities of the single particle
states of both fragments. But,
in order to have integer  numbers of nucleons in each of the two partners,
at least two Fermi energies are required. 
It is questionable when and how these
two Fermi energies are created and how the nucleus shares
the nucleons to obtain the final mass numbers. In order to understand this phenomenon,
the gaps and the Fermi energies
will be investigated at scission with the pairing density-dependent delta interaction (DDDI)
formalism and with the Woods-Saxon two center shell model.
The DDDI approach allows to determine a state-dependent pairing interaction while
the two center shell model offers the possibility to identify the localization
of the single particle states in the fission fragments.
For reflection-symmetric fission, the distribution of single particle occupation probabilities
is the same for the two similar single particle
levels schemes of the fragments. Therefore,
the numbers of nucleons should be the same in the two fission products.
The problem arises when reflection-asymmetries are 
considered at scission. The BCS theory with a constant pairing force allows only one
distribution of single particle occupation probabilities that depends on the
whole nuclear structure. 
This distribution of occupation probabilities is constrained
by the total number of nucleons and
gives non-integer values of the nucleon numbers in the two partners. 
Therefore, the alpha-decay of a superheavy element considered as a superasymmetric
fission process will be treated as a limiting case.

\section{The macroscopic-microscopic approach in the two center shell model}

Microscopic approaches based on the Hartree-Fock
theory or the macroscopic-microscopic method have been used in the
investigation in disintegration processes.
The former approach is the more fundamental way to determine
the driving potential. It starts with a realistic force between nucleons
and constructs an appropriate many-body equation. 
Surpassing huge numerical difficulties, this approach can be used
now to determine the barrier in cluster decay \cite{warda}
or for fission \cite{umar,lu}. However, the way in which the passage
from one nucleus into two separated bodies is envisaged remains a peculiar
behavior of this picture. The basic
idea of the alternative more phenomenological approach is that a
macroscopic model, as the liquid drop one, describes quantitatively
the smooth trends of the potential energy with respect to the
particle number and the deformation whereas a microscopic formalism
such as the shell model describes local fluctuations. The
mixed macroscopic-microscopic method should reproduce both smooth
trends and local fluctuations. The reliability of the latter approach was
 already tested in the case of $\alpha$-decay by considering the process
like a superasymmetric spontaneous fission \cite{epl} process. For the microscopic part
the Woods-Saxon two center shell model is used \cite{prc78}. 

In the macroscopic-microscopic method, the whole system is characterized by some collective coordinates that
determine approximately the behavior of many other intrinsic variables \cite{ni,swia,deni}. 
The basic ingredient in such an analysis
is the shape parametrization that depends on several macroscopic degrees of freedom. The generalized coordinates
associated with these degrees of freedom vary in time leading to a split of the nuclear system in two separated
fragments. The model is valid as long as the time-dependent variations
of the generalized coordinates make sense.

\subsection{Nuclear shape parametrization}

\begin{figure}
\resizebox{0.40\textwidth}{!}{
  \includegraphics{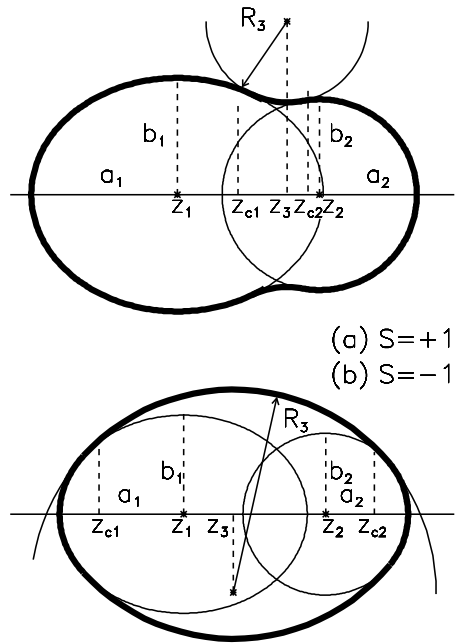}}
\caption{Nuclear shape parametrization.
}
\label{forme}
\end{figure}

We use an axial symmetric nuclear shape that offers the possibility to obtain a
continuous transition from one initial nucleus to the separated fragments. This parametrization
is obtained by smoothly joining two spheroids of semi-axis $a_i$ and $b_i$ ($i$=1,2) with
a neck surface generated by the rotation of a circle of radius $R_3$ around the axis of symmetry.
By imposing the condition of volume conservation we are left with five independent
generalized coordinates \{$q_i$\} ($i$=1,5) that can be associated to five degrees of
freedom: the elongation $R=z_2-z_1$ given by the distance between the centers of the spheroids;
the necking parameter $C_3 = S/R_3$ related to the curvature of the neck, the eccentricities $\epsilon_i=\sqrt{1-(b_i/a_i)^2}$  ($i$=1,2) associated with the deformations of the nascent fragments and the mass
asymmetry parameter $\eta = a_1 /a_2$. Alternatively, the mass asymmetry can be characterized
also by the mass number of the light fragment $A_2$. This number is obtained
by considering that the sum of the volumes of two virtual ellipsoids characterized by
the mass asymmetry parameter $\eta$ and the eccentricities $\epsilon_i$ ($i$=1,2) gives the volume
of the parent. 
The nuclear shape
is displayed in Fig. \ref{forme} where the geometrical parameters could be identified.

\subsection{Pairing corrections}

The macroscopic deformation energy is
calculated in the framework of the finite range liquid drop
model \cite{mol}.
A so called correction to the macroscopic value
is then evaluated using the Strutinsky procedure \cite{hills}.
The Strutinsky effects contain two terms: a shell  correction
and a pairing one. The shell corrections take into account
non-uniformities in the nuclear structure as function of the
deformation or the number of nucleons. No information about
the pairing are given by the macroscopic energy or the shell effects.
Therefore, only the pairing corrections are interesting for our 
treatment.

The pairing corrections in the framework of the macroscopic-microscopic model
are given by the difference
\begin{equation}
\delta P=E_p-\tilde{E}_p
\label{sco}
\end{equation}
between an exact value and an averaged one.
The total energy with paring interactions can be approximated with the relation
\begin{equation}                     
E_p=2\sum_k\rho_k\epsilon_k-\sum_k u_kv_k\sum_{k'}u_{k'}v_{k'}G_{kk'}-\sum_k
G_{kk}v_k^4
\end{equation}
The energy of a system without taking into consideration the
variations of occupation probabilities  and for an uniform
distribution of levels reads \cite{pomor}
\begin{equation}
\tilde{E}_p=2\sum_{k\le k_F}\epsilon_k-{1\over 2}\tilde{g}(\tilde{\lambda})
\tilde\Delta^2-G{N\over 2}
\end{equation}
where $k_F$ labels the Fermi level, $\tilde{g}(\tilde{\lambda})$ is the average
level density at the smoothed Fermi energy $\tilde{\lambda}$, $N$ the number 
of nucleons, $G$ is the averaged pairing interaction, and
$\tilde\Delta=12/A^{1/2}$ MeV is
an average gap \cite{hills,pomor}.
It is assumed that the last two terms in the average pairing energy
\begin{equation}
E_a=-{1\over 2}\tilde{g}(\tilde{\lambda})
\tilde\Delta^2-G{N\over 2}
\label{emedie}
\end{equation}
with the interaction strength
\begin{equation}
{1\over G}=\tilde{g}(\tilde{\lambda})\ln\left({2\Omega\over 2}\right)
\end{equation}
is contained already in the liquid drop part of the total energy \cite{myers}.
Therefore, a problem is raised by this average pairing energy 
when the nucleus is split into two fragments. The authors of Ref. \cite{pomor}
emphasized the fact that a double counting should be envisaged 
at scission that can be eliminated by considering   
a dependence of the pairing strength with deformation \cite{poma}.

\subsection{The Woods-Saxon two center shell model}

The many-body wave function and the single particle energies
are provided  by the  Woods-Saxon two-center shell model \cite{prc78}. The Woods-Saxon
potential, the Coulomb interaction and the spin orbit term must be diagonalized in a
double center eigenvectors basis.
A complete analytical eigenvectors basis can be only obtained
for the semi-symmetric two-center oscillator. This
potential corresponds to a shape parametrization given
by two ellipsoids that possess the same semi-axis perpendicular
on the axis of symmetry. The potential
is
\begin{equation}
V_{o}(\rho,z)=\left\{\begin{array}{lc}
{1\over2}m\omega_{z1}^{2}(z-c_{1})^{2}+{1\over2}m\omega_{\rho}^{2},& z<0,\\
{1\over2}m\omega_{z2}^{2}(z-c_{2})^{2}+{1\over2}m\omega_{\rho}^{2},& z\ge0,
\end{array}\right.
\end{equation}
where $\omega$ denotes the stiffness of the potential along
different directions as follows, $\omega_{z1}=\omega_{0}{R_{0}\over a_{1}}$,
$\omega_{z2}=\omega_{0}{R_{0}\over a_{2}}$, $\omega_{\rho}=\omega_{0}{R_{0}\over b_{1}}$,
$\omega_{0}=41A_{0}^{-1/3}$, $R_{0}=r_{0}A_{0}^{1/3}$, in order to
ensure a constant value of the potential on the surface. The origin
on the $z$-axis is considered as the location of the plane of
intersection between the two ellipsoids.

The asymmetric  two center shell oscillator provides an orthogonal eigenvectors basis for only one Hermite space
\cite{prc96,npa2c}.
An analytic system of eigenvectors can be
obtained for $V_{0}$ by solving the   Schr\"{o}dinger equation:
\begin{equation}
\left[-{\hbar^{2}\over 2m_{0}}\Delta+V_{o}(\rho,z)\right]\Psi(\rho,z,\varphi)=E\Psi(\rho,z,\varphi)
\label{eqsp2}
\end{equation}
The analytic solution of Eq. (\ref{eqsp2}) is obtained using the ansatz
\begin{equation}
\Psi(\rho,z,\varphi)=Z(z)R(\rho)\Phi(\varphi)
\label{eivb}
\end{equation}
with
\begin{equation}
\Phi_{m}(\varphi)={1\over\sqrt{2\pi}}\exp(im\varphi)
\end{equation}
\begin{equation}
R_{nm}(\rho)=\sqrt{2n!\over (n+m)!}\alpha_{\rho}\exp\left(-{\alpha_{\rho}^{2}\rho^{2}\over 2}\right)
(\alpha_{\rho}\rho)^{m}L_{n}^{m}(\alpha_{\rho}^{2}\rho^{2})
\end{equation}
\begin{equation}
Z_{\nu}(z)=\left\{ \begin{array}{l}
C_{\nu_{1}}\exp\left(-{\alpha_{z1}^{2}(z-c_{1})^{2}\over 2}\right){\bf{H}}_{\nu_{1}}[-\alpha_{z1}(z+c_{1})],\\~~~~~~~~z<0;\\
C_{\nu_{2}}\exp\left(-{\alpha_{z2}^{2}(z-c_{2})^{2}\over 2}\right){\bf{H}}_{\nu_{2}}[\alpha_{z2}(z-c_{2})],\\~~~~~~~z\ge 0,
\end{array}\right.
\label{basis}
\end{equation}
where $L_{n}^{m}(x)$ is the Laguerre polynomial, ${\bf{H}}_{\nu}(\zeta)$ is the Hermite function,
$\alpha_{l}=(m_{0}\omega_{l}/\hbar)^{1/2}$ ($l=z_1,z_2,\rho$) are length parameters,
and $C_{\nu_{i}}$ ($i=1,2$) denote the normalization constants.
The quantum numbers $n$ and $m$ are integers while the quantum numbers $\nu_i$ along the $z$-axis are real
and have different values for the intervals $(-\infty,0]$ and $[0,\infty)$.
Imposing conditions for the continuity of the wave function and its derivative, together
with those for the stationary energy and orthonormality, the values of $\nu_{1}$, $\nu_{2}$, and of the normalization constants $C_{\nu_{1}}$, and
$C_{\nu_{2}}$ could be obtained.
Details concerning these solutions and expressions for the normalization constants
are found in Refs. \cite{prc96,npa2c}.
For reflection-symmetric shapes, the solutions along the $z$-axis are also characterized
by the parity as a good quantum number.
The basis (\ref{basis}) for the two-center
oscillators can be used for  various ranges of models which are
more or less phenomenological ones \cite{maruhn,geng,torres,torres2,neste,sun,hass}.

In this unique Hermite space, the behavior of both fragments 
can be described simultaneously.
The orthogonal wave functions are centered in one of the two regions in three-dimensional  position space.
Each wave function is analytically continued in both regions.
In an intermediate situation of two partially overlapped potentials, 
each eigenfunction has components in the two subspaces that belong to the fragments.
When the elongation $R$ is zero, the eigenvectors
basis becomes that of a single anisotropic oscillator and the Hermite function is transformed
into an Hermite polynomial. When $R$ tends to infinity, a two oscillators
eigenvectors system is obtained naturally in the same Hermite space,
centered in the middle of the two fragments. Asymptotically, when
the elongation tends to infinity, only one of the
values $\nu_1$ or $\nu_2$ is transformed into an integer. The associated
Hermite function becomes a Hermite polynomial with its proper normalization
constant. In the same time, the normalization constant of the other Hermite
function reaches a zero value. The Pauli principle is fulfilled. Even for
symmetric two center potentials, only one state is simultaneously active
in all the position space.
Therefore, the two center shell model always provides the wave functions associated
to the lower energies of the single particle states pertaining to a major quantum number $N_{max}$.
As a consequence, molecular states formed by two fragments at scission could be precisely described.
In contrast to the cluster approximations,
the two-center shell model offers the opportunity to treat fission in a wide
range of mass asymmetries \cite{extr,rrp} and offers to opportunity 
to consider the alpha decay as a 
superasymmetric fission process \cite{epl,mal,rjp}.

If the wave function is located in only one of the two potential wells, it is
possible to identify the single particle states that belong to each fragment
at scission.  The square of the single particle wave function of each state is integrated
separately in the two subspaces and two probabilities are obtained. 
The single particle wave function
is located in the well characterized by the larger probability. Such tasks
were already performed \cite{m1,m2,m3} in order to estimate
the partition of the excitation energy in fission processes between the two partners.

\section{The density-dependent delta interaction}

A simple treatment of the pairing interaction in which the major
contribution to the residual interaction is coming from the nuclear surface
region is given by the
DDDI model \cite{rrc}. The following spatial modulation   
of the pairing strength is postulated \cite{doba,bender,yoshida, tajima} as follows:
\begin{equation}
V_p(\vec{r})=-V_{0p}\left[1-\beta\left({\rho(\vec{r})\over \rho_0}\right)^\gamma\right]
\label{i0}
\end{equation}
The fitting parameters $\beta$=1 and $\gamma$=1 are usually considered as unity. 
The saturation density of the
nuclear matter is $\rho_0$=0.16 fm$^{3}$, while the $\rho(r)$ is
the local density.
We consider here that the local nuclear density is shape dependent and its behavior
can be described by a Woods-Saxon function, in accordance to the shape of the
potential used.
The parameters $V_{0p}$ of the interaction (\ref{i0}) should be adjusted 
to reproduce observables.
In our case, the observable is the pairing gap.
Values of the model dependent strength parameters $V_{0p}$ are
deduced in several works, as for example in Ref. \cite{bender},
 $V_{0p}$ being -999 MeV fm$^{-3}$ and -1146 MeV fm$^{-3}$ for neutrons and protons, respectively.
The values deduced in our treatment are different, as it will be specified later.
The calculations based on this schematic pairing interaction 
could be solved on a finite space
of states, called active
levels pairing space. This space is limited by some cutoff in the single particle energies.

For the pairing field given by formula (\ref{i0}), the state-dependent
 pairing interaction strengths $G_{ij}$ are defined as:    
\begin{equation}
G_{ij}=-\int V_p(\vec{r})\mid \phi_i(\vec{r})\mid^2 \mid\phi_j(\vec{r})\mid^2 d\vec{r}
\end{equation}
where $ \phi_j(\vec{r})$ is the wave function of the state $j$ in
the position representation.
The pairing  state-dependent gaps resort from the BCS equations
\begin{equation}
\Delta_i={1\over 2}\sum_j{G_{ij}\Delta_j\over \sqrt{(\epsilon_j-\lambda)^2+\Delta_j^2}}
\label{pairg}
\end{equation}
that are correlated by the condition of a fixed number of pairs $N_p$
in the active levels pairing 
space:
\begin{equation}N_p=\sum_i v_i^2. \end{equation}
In the previous equalities, it is considered that the index $i$ run over 
$2N_p$ levels around the Fermi energy. 
 The occupation
probability of the state $i$ is
\begin{equation}
v_i^2={1\over 2}\left[1-{\epsilon_i-\lambda\over\sqrt{(\epsilon_i-\lambda)^2+\Delta_i^2}}\right]
\end{equation}
while the vacancy probability is $u_i^2=1-v_i^2$. In the previous equation,
 $\lambda$ represents the Fermi
energy of the whole nuclear system. 
The averaged pairing energy gap of the whole system could be considered as 
\begin{equation}
\bar\Delta={\sum_i u_iv_i\Delta_i\over\sum_i u_iv_i}
\label{gapm}
\end{equation}
This definition emphasizes the importance of the
single particle levels located in the vicinity of the
Fermi energy because the weights $u_iv_i$ are larger in this region.

\begin{figure}
\resizebox{0.40\textwidth}{!}{
  \includegraphics{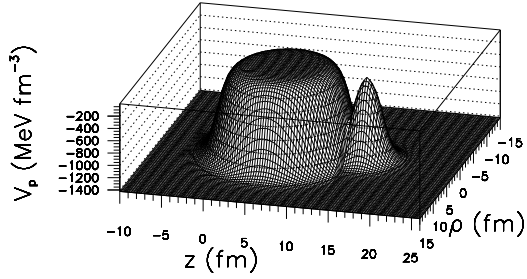}}
\caption{The pairing strength $V_p$  as function of the cylindrical
coordinates $\rho$ and $z$ that characterize the shape
of the nuclear system at a distance between the centers of
the fragments $R$= 10 fm.
}
\label{echip}
\end{figure}

 In order to determine the parameter $V_{0p}$ of the
pairing field for a given levels pairing space, we must refer to the empirical
 values of the pairing gaps. 
  A reliable value for the energy gap can be 
obtained with the five points formula \cite{maruhn}
that takes into account the experimental masses. 
Unfortunately, no enough experimental or evaluated masses are available to date
\cite{audi} in the vicinities of the nuclei $^{296}$Lv and $^{292}$Fl.
It is not possible 
to extract reliable empirical pairing gap values. Therefore,
our estimations rely on theoretical evaluations.
The theoretical tables \cite{moller} give the next values for $^{296}$Lv:
$\Delta_p$=0.66 MeV and $\Delta_n$=0.56 MeV, where the index $p$ or $n$ denotes  
proton or neutron, respectively. For $^{292}$Fl we have $\Delta_p$=0.71 MeV
and $\Delta_n$=0.60 MeV.
The $^{296}$Lv theoretical interaction strengths
$V_{0p}$ for the pairing matrix elements that are able to reproduce these gap values
were estimated as -1448 MeV fm$^{-3}$ and -1528 MeV fm$^{-3}$ 
for neutrons and protons,
respectively. In the case of the $^{292}$Fl, we obtained -1440 MeV fm$^{-3}$
for neutrons and -1605 MeV fm$^{-3}$ for protons.  
We used two active levels pairing spaces of $2N_p$=190 and 
$2N_p$=116 single particle levels,
corresponding to the $^{296}$Lv numbers of neutrons and protons, respectively.
It is considered that the values of the pairing strengths vary linearly from
the ground state of the parent nucleus up to the scission configuration. In 
other words, the nascent $\alpha$ particle feels the pairing field of the daughter. 
The pairing strength spatial modulation is represented in the Fig. \ref{echip} for
the touching configuration between the daughter nucleus and the $\alpha$ particle.

\begin{figure}
%\begin{center}
\resizebox{0.50\textwidth}{!}{
  \includegraphics{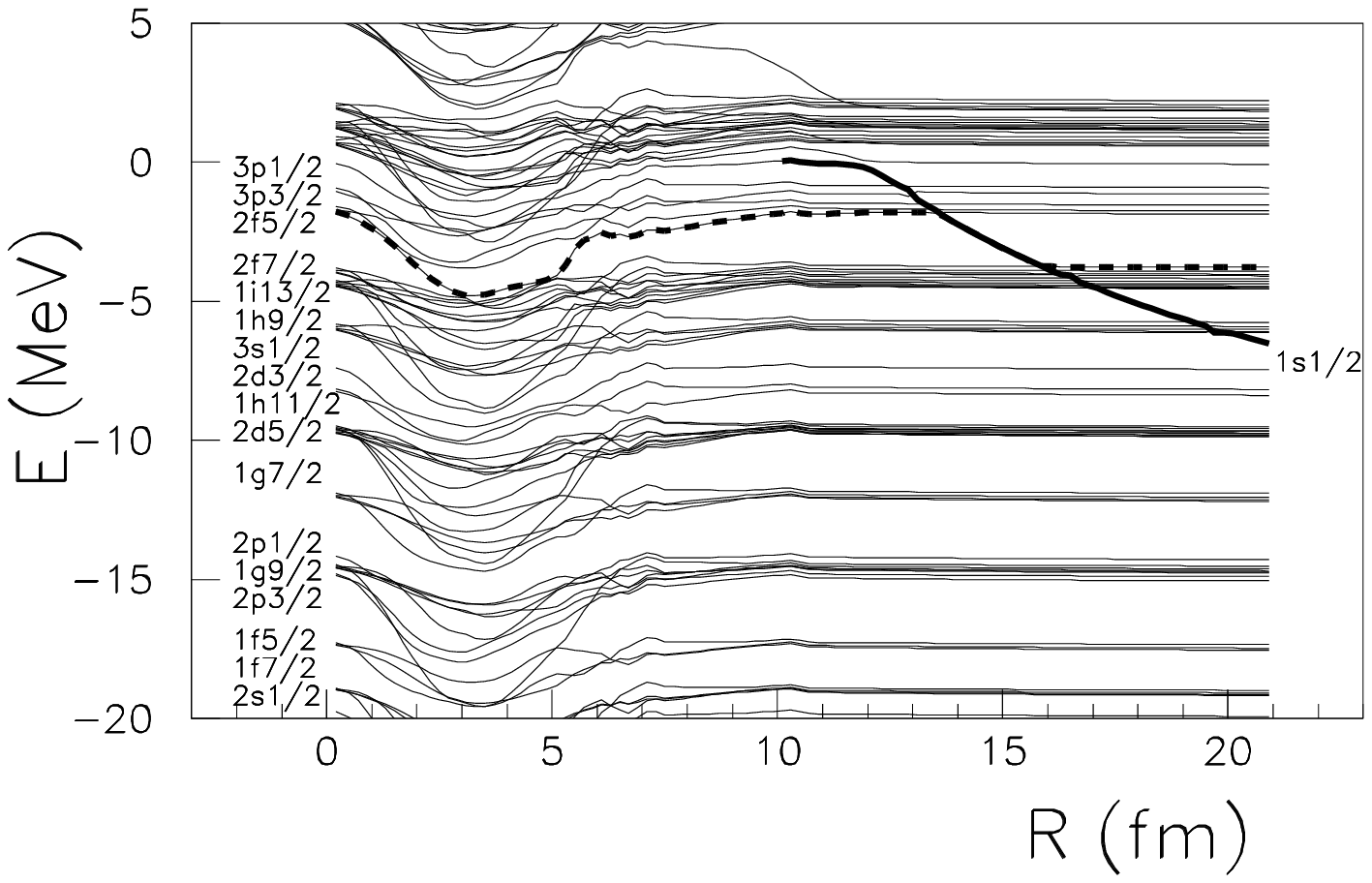}}
\caption{Single particle level scheme for protons in the region of the Fermi energy
as function of the elongation $R$ for $\alpha$-decay of $^{296}$Lv. 
The Fermi level is plotted with a thick dashed line
while the level pertaining to the alpha nucleus is plotted with thick solid
lines.
The orbital of the alpha nucleus is marked on the right.
}
\label{ttp}
\end{figure}

\begin{figure}
%\begin{center}
\resizebox{0.50\textwidth}{!}{
  \includegraphics{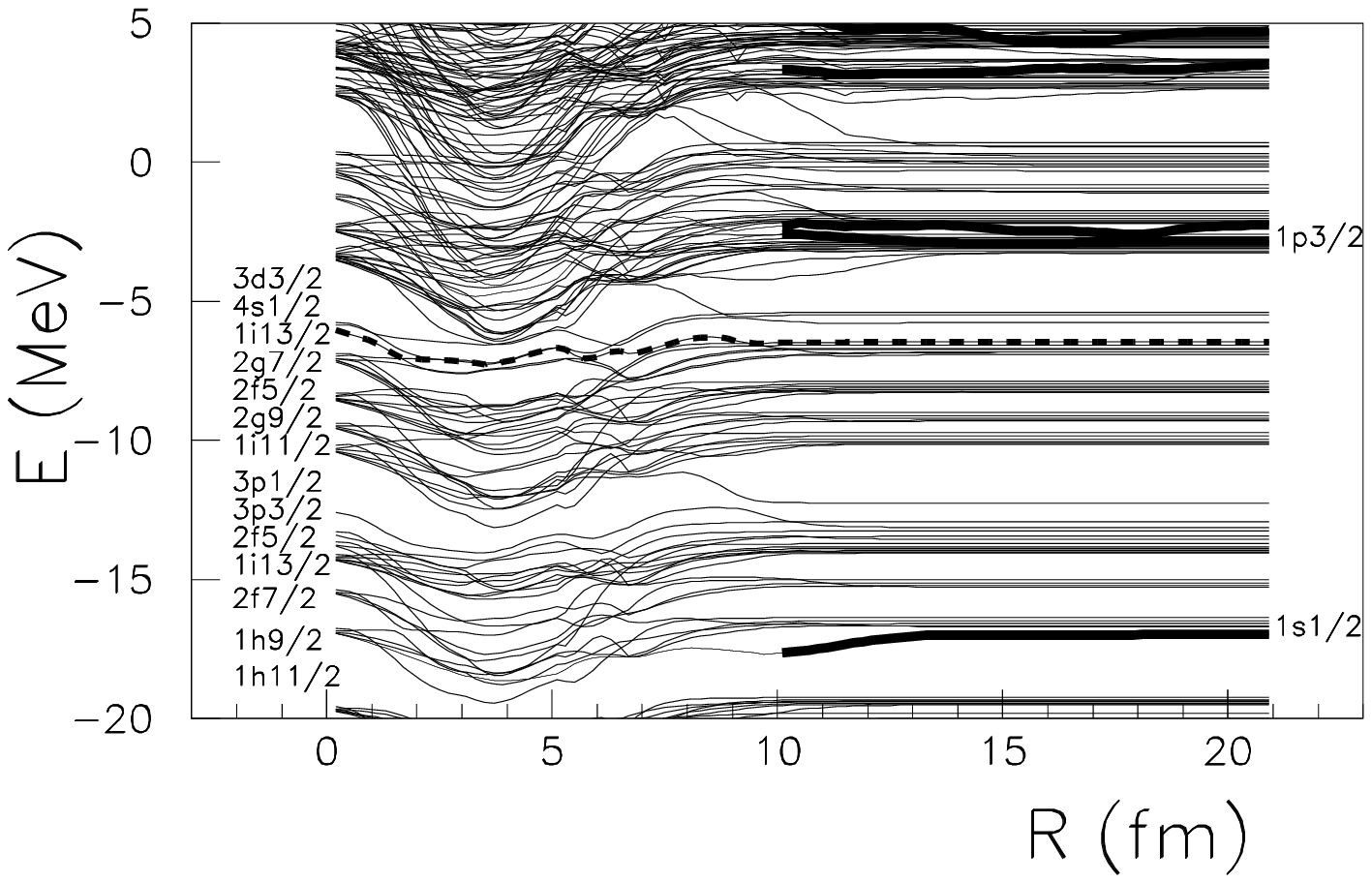}}
\caption{Single particle level scheme for neutrons in the region of the Fermi energy
as function of the elongation $R$ for $\alpha$-decay of $^{296}$Lv. 
The Fermi level is plotted with a thick dashed line
while the levels pertaining to the alpha nucleus are plotted with thick solid
lines.
The orbitals of the alpha nucleus are marked on the right.
}
\label{ttn}
\end{figure}

\begin{figure}
\resizebox{0.50\textwidth}{!}{
  \includegraphics{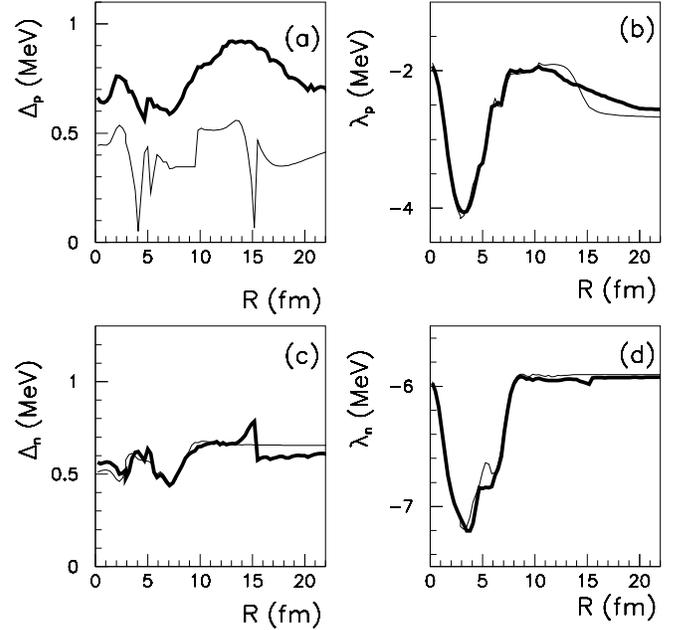}}
\caption{Energy pairing gaps and Fermi energies. (a) The proton pairing gap $\Delta_p$
in the DDDI approach as function of the elongation $R$ is displayed
with a full thick line. The proton pairing gap for a constant
strength value $G$ is plotted with a thin line.
(b) The proton Fermi energy $\lambda_p$ of the whole nuclear
system in the DDDI approach is plotted with a full thick line as function
of the elongation $R$. The Fermi energy in the constant $G$ approximation
is displayed with a thin line.
(c) The neutron pairing gap $\Delta_n$
in the DDDI approach as function of the elongation $R$ is displayed
with a full thick line. The proton paring gap for a constant
strength value $G$ is plotted with a thin line.
(d) The neutron Fermi energy $\lambda_n$ of the whole nuclear
system in the DDDI approach is plotted with a full thick line as function
of the elongation $R$. The Fermi energy in the constant $G$ approximation
is displayed with a thin line.
}
\label{gaptotal}
\end{figure}

\begin{figure}
%\begin{center}
\resizebox{0.50\textwidth}{!}{
  \includegraphics{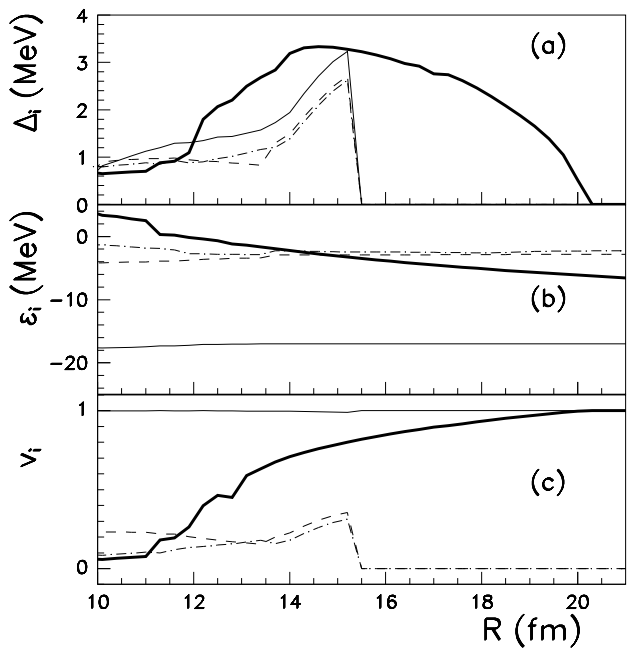}}
\caption{(a) Values of the state-dependent pairing gap parameters $\Delta_i$ for the lower
single-particles states that reach
the $\alpha$ nucleus after the scission as function
of the elongation $R$. The thick full
line corresponds to the proton asymptotic state $1s_{1/2}$, 
the thin full line is for
the neutron final state $1s_{1/2}$, while the dot dashed and dashed lines
are associated to the two $1p_{3/2}$ neutron states.
(b) Single particle energies $\epsilon_i$ for these orbital. 
The same line types as in the caption (a) identify the states.
(c) BCS amplitudes $v_i$ for the above mentioned states. The 
line types refer to the same states as in caption (a).
}
\label{alfagap}
\end{figure}
\begin{figure}
%\begin{center}
\resizebox{0.50\textwidth}{!}{
  \includegraphics{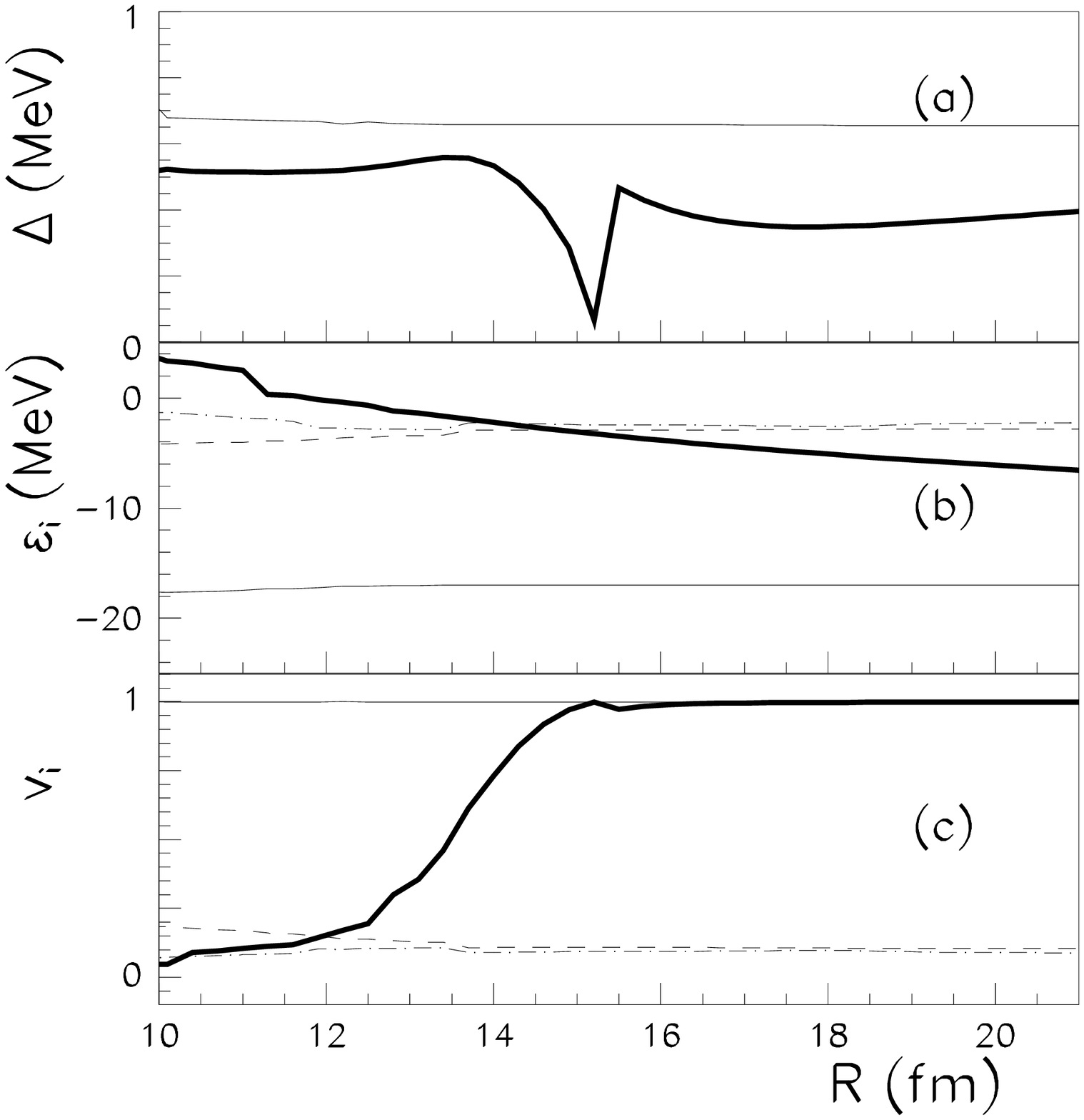}}
\caption{(a) Values of the pairing gap parameters $\Delta$ in the BCS approximation
with a constant  value of the pairing interaction $G$. The thick line represents
the pairing gap for proton while the thin line indicates the pairing gap for neutron. 
(b) Single particle energies $\epsilon_i$ for the orbitals that reach
the alpha nucleus. The same line types as in Fig. \ref{alfagap} are used.
(c) BCS amplitudes $v_i$
calculated with constant pairing interaction approach. 
The line types refer to the same 
states as in caption (b).}
\label{alfagapc}
\end{figure}

\section{Results}

In our previous study, a path for the superasymmetric fission
leading to alpha emission was obtained for $^{296}$Lv \cite{epl}. This
least action path is assigned along a so called $\alpha$ valley in
the deformation energy surface
and crosses a molecular minimum. The $\alpha$ particle
is born in this molecular minimum.  
The scission configuration is produced for
an elongation $R\approx$ 10 fm.

The single particle levels schemes along the
superasymmetric fission path are displayed in Figs. \ref{ttp} and \ref{ttn},
for protons and neutrons, respectively. An energy window
around the Fermi level is selected to evidence the mean features
 of the process. As mentioned, close to the scission configuration
the two center shell model offers the possibility 
to identify the single particle levels that pertain to the 
alpha particle. These levels are plotted with thick solid lines.
Asymptotically, the model gives a superposition of both the single particle level schemes of
the daughter nucleus and of the $\alpha$ particle. If the BCS equations
with constant pairing strength $G$ are solved in the region of separated nuclei,
this superposition of single particle level schemes should be taken into
consideration, no matter the affiliation of the nucleons in one of the nuclear fragments.
The Fermi
single particle levels are plotted with a thick dashed line in both
figures. A polarization effect is clearly visible for the
1s$_{1/2}$ proton single particle level associated to the $\alpha$ particle.
The energy of this level decreases when the two nuclei get away
one from another. After the scission, this level crosses many shells of the daughter nucleus.
The proton energies of the daughter nucleus decrease also after the scission, but
with a much smaller slope.

First of all, the Fermi energies and the pairing gaps of the whole
nuclear system are calculated along the superasymmetric fission trajectory
in the framework of the DDDI approach and (for comparison) under the
hypothesis of a constant pairing strength $G$. 
We calculated the averaged pairing gaps of the whole
nuclear system with the Eq. (\ref{gapm}).
The results are displayed
in Fig. \ref{gaptotal}. A thick line is used for the
DDDI formalism while a thin one for the $G$ constant approximation.
Both approaches give a gross similar structure 
in the variation of the pairing gaps and of the Fermi energies
as function of the distance between the centers of the fragments $R$.
The values of the DDDI proton pairing $\Delta_p$ are always
larger than those given by the constant $G$ approximation. A sudden
variation is obtained for the DDDI neutron pairing gap $\Delta_n$
at $R\approx$ 15 fm. This variation could be understood 
in the following analysis of the
state dependent pairing gaps. It is important to note that the
average values of the gaps never exceed 1 MeV for both isospins.

In Fig. \ref{alfagap}(a), the pairing gaps $\Delta_i$ associated to the 
single particle states that pertain to the $\alpha$ potential
well after the scission are plotted as function of the 
elongation $R$. For simplicity, only single particle levels with energies located
in the vicinity  
of the Fermi energy of the compound system are displayed. In the case of proton, only
the $\alpha$ 1s$_{1/2}$ state could be selected in the given energy window. 
In the case of neutron, three single particle levels
are identified. 
The energies of all these single particle levels are
 displayed with thick lines for proton and thin lines for neutron in Fig. \ref{alfagap}(b).
In the Fig. \ref{alfagap}(c), the occupation
amplitudes $v_i$ of the selected single particle states are plotted.  
The gap associated to the neutron 1$s_{1/2}$ $\alpha$-state
increases  up to a value
close to 3.5 MeV at an elongation $R\approx$ 14 fm. 
After this distance between the centers of the
nascent fragments, the
gap drops to zero but the occupation amplitude reaches
the value of unity. The gaps associated to the superior
neutron single particle levels have a different behavior. Their
gaps drop to zero at the same value of the elongation while
the occupation amplitudes reach  zero values.
In the case of the proton 1$s_{1/2}$ single particle state, 
the gap reaches a zero value at an elongation close to 19 fm
and its occupation amplitudes becomes unity. These behavior
reflect the short range character of the nuclear forces and
the long range character of the Coulomb interaction.
After a certain distance between the two nuclei
($R\approx$ 10-19 fm), the pairing interactions $G_{ij}$
between orbitals located in different single particle potentials become zero. 
The connections between the pairing gaps $\Delta_{i}$ for single particle states
located in different fragments given by
the values of $G_{ij}$ in Eqs. (\ref{pairg}) are lost.
Therefore, the pairing equations (\ref{pairg}) are nearly transformed into
two separated systems of equations, one for the daughter and another for
the alpha particle, correlated only by the
value  of the Fermi energy $\lambda$. 
It was already noticed that the average values of the DDDI pairing gaps
never exceed 1 MeV. In consequence, it can be assessed that
the values of the pairing gaps associated to the
alpha particle are always larger than those associated to
the daughter nucleus in the vicinity  of the scission configuration.
These large values of the $\alpha$
energy gaps close to 3.5 MeV at scission have a particular importance
for the alpha decay process. The quasiparticle energies of the
alpha states are very large. 
That means,  the quasiparticle excitations of 
the nascent alpha particle close to scission  are strongly suppressed.

For comparison purposes, the occupation probabilities of the
single particle levels were also calculated in the constant pairing
interaction approach. The value of the interaction $G$ is obtained
within a normalization procedure that takes into account the density
of levels at the Fermi energy and the average gap $\tilde{\Delta}$ \cite{hills}.
The results are plotted in Fig. \ref{alfagapc}.
The values of the gap $\Delta$ as function of the internuclear
distance are displayed in the panel (a). The values of the
gap parameters never reach a zero value, being the same as
for all nucleons of the same specie of the whole nuclear system. 
A fluctuation of the proton pairing gap can be observed at $R\approx$ 15.2 fm.
This fluctuation reflects the crossing between the 1s$_{1/2}$ single
particle state of the $\alpha$ particle with a shell of the
daughter located in the vicinity of the Fermi energy. The proton
single particle level can be identified in panel (b).
In panel (c) the occupation amplitudes are plotted.
The amplitude of the proton state reaches the value of unity
at an internuclear elongation smaller than that in the case obtained with
DDDI. The lowest neutron state 1s$_{1/2}$ has practically a constant 
occupation amplitude, slightly lower than unity. 
The outer neutron orbitals (1$p_{3/2}$) remain with 
non-zero values of their occupation amplitudes, even after the scission.
That means, the emitted nucleus has a (real) number of neutrons larger that
2 in the constant $G$ approximation.
 
A Fermi energy $\lambda'$for the daughter can be obtained by solving
the equation for the number of particles
\begin{equation}N_{d}=\sum_{i'} v_{i'}^2. \end{equation}
where it is considered that the index $i'$ runs over 
$2N_{dp}$ levels for states located in the daughter well
that can be provided by the active level pairing space. 
2$N_d$ is the number of nucleons of the daughter. 
 The occupation
probability of the state $i'$ is
\begin{equation}
v_{i'}^2={1\over 2}\left[1-{\epsilon_{i'}-\lambda'\over\sqrt{(\epsilon_{i'}-\lambda')^2+\Delta_{i'}^2}}\right]
\end{equation}
Consequently, $\lambda'$ is now the Fermi energy of the daughter.

\begin{figure}
%\begin{center}
\resizebox{0.50\textwidth}{!}{
  \includegraphics{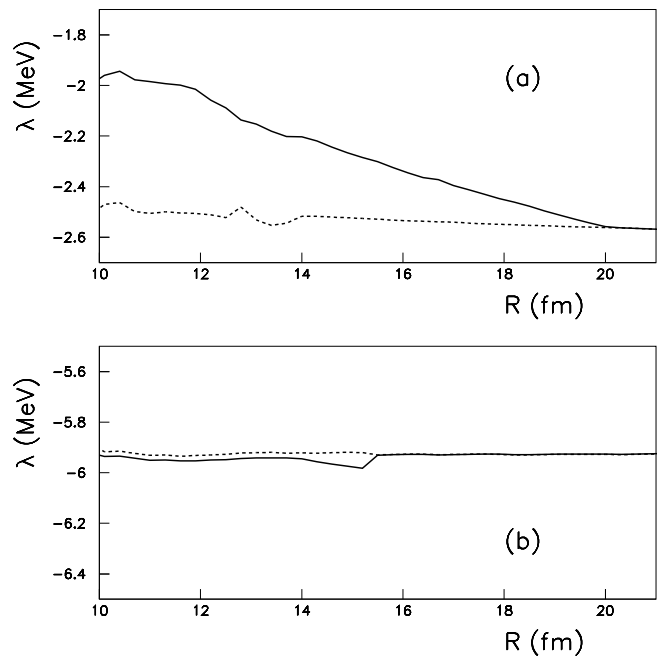}}
\caption{The whole system Fermi energies $\lambda$ for proton (a) and neutron (b)
are plotted with full lines. The Fermi energies of the daughter $\lambda'$
are displayed with a dashed line.
}
\label{alfafermi}
\end{figure}

\begin{figure}
%\begin{center}
\resizebox{0.50\textwidth}{!}{
  \includegraphics{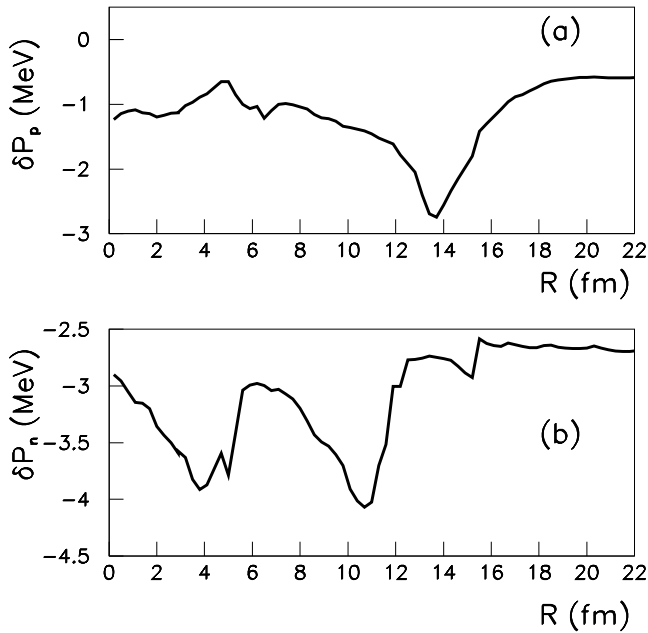}}
\caption{Pairing correction $\delta P$ for proton (a) and neutron (b) computed with a
state dependent pairing interaction $G_{ij}$.
}
\label{corpat}
\end{figure}

\begin{figure}
%\begin{center}
\resizebox{0.50\textwidth}{!}{
  \includegraphics{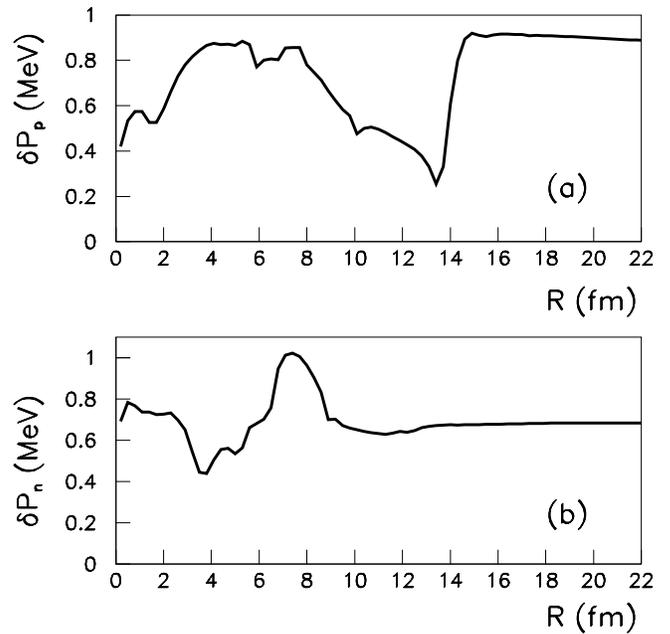}}
\caption{Pairing correction $\delta P$ for proton (a) and neutron (b) computed with a
constant value of the interaction $G$.
}
\label{alefpc}
\end{figure}

In Fig. \ref{alfafermi}, the Fermi energies $\lambda$ of the whole system
are plotted with a full line while the dashed lines correspond
to the Fermi energies of the daughter $\lambda'$. The panel (a) corresponds
to proton while the (b) one to neutron. It is interesting to note
that $\lambda$ becomes equal to $\lambda'$ asymptotically, no matter
the isospin.

In Fig. \ref{corpat}, the pairing corrections for proton $\delta P_p$ (a) and for 
neutron $\delta P_n$ (b) are
calculated with the formula (\ref{sco}) 
along the superasymmetric fission
trajectory in the DDDI approach. 
The pairing correction formalism is not an appropriate treatment
in the case of the $\alpha$ particle. Terms like those
given by the expression (\ref{emedie}) cannot be constructed.
Therefore, the terms proportional
to $G_{ii}$ in which $i$ denote states of the $\alpha$ particle
were gradually suppressed after the scission in formula (\ref{sco}). Asymptotically,
we are left with the pairing effects of the daughter.
The values of the proton
pairing correction $\delta P_p$ exhibit a  structure with a negative fluctuation
at $R\approx$ 13-14 fm. 
Usually the pairing corrections increase for low density
of levels around the Fermi energy. 
So, this negative fluctuation reflects the fact
that the single particle energy of the $\alpha$ 1s$_{1/2}$ state crosses
the Fermi energy of the whole system, as displayed in Fig. \ref{ttp}.
A structure can be observed also in the case of the 
neutron pairing corrections $\delta P_n$. Close to $R\approx$ 15 fm,
a rapid fluctuation is produced. This fluctuation corresponds exactly
to the moment in which the pairing gaps of the states that are located
in the $\alpha$ nucleus become zero, as it can be seen in the Fig. \ref{alfagap} (a).
A peak can be also observed in Fig. \ref{corpat} (b) at $R\approx$ 8 fm.
This peak reflects that a low density of states is realized around
the Fermi energy, as exhibited in Fig. \ref{ttn}.

For comparison, the pairing corrections 
space are represented in Fig. \ref{alefpc}
with the constant interaction $G$ approach.
The structure of the pairing corrections resemble to that given
by the state dependent pairing interactions method. However, the fluctuations
at $R\approx$ 15 fm obtained for neutron are missing.

\section{conclusion}
The alpha decay was treated as a superasymmetric fission process
with particular emphasis on the scission configuration.
The pairing interaction was taken into account through the DDDI  approach
in terms of the two center shell model.
The pairing gaps of the single particle states pertaining to  the nascent
alpha particle reach very large values close to scission. These
values of the pairing gaps certainly suppress the quasiparticle excitation
probabilities. So, the alpha particle forms a strong bound cluster when escaping
from the parent nucleus. We have a mechanism in which the $\alpha$ particle
can be considered preformed on the surface of the daughter nucleus.
When the overlaps between the wave functions that pertain
to different fragments  vanish, the pairing gaps of the alpha
particle become zero. In the same time, the Fermi energy of the daughter
nucleus reaches the value evaluated for the compound nuclear system.
The projection of particle numbers was obtained in a natural way.

Many models invoke a preformation of the alpha particle in the parent nucleus
in order to penetrate an Coulomb external barrier \cite{av1,sil1,san1,lov1}.
Such preformations are invoked also for ternary fission \cite{car1}.
The preformation is proportional
to the square overlaps between the ground state wave functions of the parent
and the antisymmetric product between the wave functions of the nascent
fragments
in different configurations after the scission \cite{mang,san2}. Our
calculations
show that when the nuclear shape parametrization describes very
asymmetric systems, the nuclear matter builds  a very bound cluster
consisting of two protons and two neutrons on its surface. This cluster
survives in the mean field created by the remaining nucleons.
It was also assessed, at least formally, that the preformation
of an emitted particle and its barrier penetrability between
the ground state and the scission point are quantities
with the same significance \cite{po,zd}. This equivalence gives a support
in the attempt to investigate the $\alpha$ decay process using fission
theories in order to understand the mechanism of its formation.
The DDDI formalism in conjunction with the two center shell model
can offer a valuable tool for the investigation of scission properties
of nuclear disintegration.

\subsection{Acknowledgements}
The author is indebted to Krzysztof Pormorski for giving notice of the
subject and for illuminating discussions. This work was supported by CNCS-UEFISCDI, Project
No. PN-II-ID-PCE-2011-3-0068.


\begin{thebibliography}{00}
\bibitem{younes}W. Younes, and D. Gogny, Phys. Rev. Lett. {\bf 107},
132501 (2011).
\bibitem{warda}W. Warda, and J.L. Egido, Phys. Rev. C {\bf 86}, 014322 (2012).
\bibitem{umar} C. Simenel, and A.S. Umar, Physical Review C {\bf 89}, 031601(R) (2014).
\bibitem{lu}B.N. Lu, E.-G. Zhao, and S.-G. Zhou, Phys. Rev. C {\bf 85},
011301 (2012).
\bibitem{epl}A. Sandulescu, M. Mirea, and D.S. Delion
EPL {\bf 101}, 62001 (2013).
\bibitem{prc78}M. Mirea, Physical Review C {\bf 78}, 044618 (2008).
\bibitem{ni}J.R. Nix, Ann. Rev. Nucl. Sci. {\bf 22}, 65 (1972).
\bibitem{swia}W.J. Swiatecki, and S. Bjornholm, Phys. Rep. {\bf 4}, 325 (1972).
\bibitem{deni}V.Yu. Denisov, Phys. Rev. C {\bf 89}, 044604 (2014).
\bibitem{mol} P. Moller, J.R. Nix, W.D. Myer, and W.J. Swiatecki, 
Atom. Data Nucl. Data Tabl. {\bf 59}, 185 (1995).
\bibitem{hills}M. Brack, J. Damgaard, A.S. Jensen, H.C. Pauli,
V.M. Strutinsky, and C.Y. Wong, Rev. Mod. Phys. {\bf 44}, 320 (1972).
\bibitem{pomor}K. Pomorski, and F. Ivanyuk, Int. J. Mod. Phys. E {\bf 18}, 900 (2009).
\bibitem{myers}W.D. Myers and, W.J. Swiatecki, Nucl. Phys. A {\bf 81}, 1 (1966).
\bibitem{poma}B. Nerlo Pomorska and, K. Pomorski, Int. J. Mod. Phys. E {\bf 16}, 328 (2007).
\bibitem{prc96}M. Mirea, Phys. Rev. C {\bf 54}, 302 (1996).
\bibitem{npa2c}M. Mirea, Nucl. Phys. A {\bf 780}, 13 (2006).
\bibitem{maruhn}J. Maruhn, and W. Greiner, Z. Phys. {\bf 251}, 431 (1972).
\bibitem{geng} L.-S. Geng, J. Meng, and T. Hiroshi, Chin. Phys. Lett. {\bf 24}, 1865
(2007).
\bibitem{torres} A. Diaz-Torres, and W. Scheid, Nucl. Phys. A757, {\bf 373} (2005).
\bibitem{torres2} A. Diaz-Torres, Phys. Rev. Lett. {\bf 101}, 122501 (2008).
\bibitem{neste}V.A. Nesterov, Phys. At. Nucl. {\bf 76}, 577 (2013).
\bibitem{sun}Q. Sun, D.-H. Shangguan, and J.-D. Bao, Chin. Phys. C {\bf 37}, 014102 (2013).
\bibitem{hass}H. Hassanabadi, E. Maghsoodi, and S. Zarrinkamar, Few-Body Syst. {\bf 53},
271 (2012).
\bibitem{extr}A. Sandulescu, and M. Mirea, Eur. Phys. J. A {\bf 50}, 110 (2014).
\bibitem{rrp}A. Sandulescu, and M. Mirea, Rom. Rep. Phys. {\bf 65}, 688 (2013).
\bibitem{mal}M. Mirea, Phys. Rev. C {\bf 63}, 034603 (2001).
\bibitem{rjp}M. Mirea, Rom. J. Phys. {\bf 60}, in print (2015).
\bibitem{m1}M. Mirea, Physical Review C {\bf 83}, 054608 (2011).
\bibitem{m2}M. Mirea, Physics Letters B {\bf 717}, 252 (2012).
\bibitem{m3}M. Mirea,  Physical Review C {\bf 89}, 034623 (2014).
\bibitem{rrc} R.R. Chasman, Phys. Rev. C {\bf 14}, 1935 (1976).
\bibitem{doba}J. Dobaczewski, W. Nazarewicz, T.R. Werner, J.F. Berger, C.R. Chinn,
and J. Decharge, Phys. Rev. C {\bf 53}, 2809 (1996).
\bibitem{bender} M. Bender, K. Rutz, P.-G. Reinhard, and J.A. Maruhn, 
Eur. Phys. J. A {\bf 8}, 59 (2000).
\bibitem{yoshida} S. Yoshida, and H. Sagawa, Phys. Rev. C {\bf 77}, 054308 (2008).
\bibitem{tajima} N. Tajima, P. Bonche, H. Flocard, P.-H. Heenen, and M.S. Weiss,
Nucl. Phys. A {\bf 551}, 434 (1993).
\bibitem{audi}M. Wang, G. Audi, A.H. Wapstra, F.G. Kondev, M. MacCormick, X. Xu, and
B. Pfeiffer, Chin. Phys. C {\bf 36}, 1603 (2012).
\bibitem{moller}P. Moller, J.R. Nix, and K.-L. Kratz, Atom. Data. Nucl. Data. Tabl. {\bf 66}, 131 (1997).
\bibitem{av1}M. Avrigeanu, A.C. Obreja, F.L. Roman,
V. Avrigeanu, and W. von Oertzen, At. Data Nucl.
Data Tabl. {\bf 95}, 501 (2009).
\bibitem{sil1}A.I. Budaca, and I. Silisteanu, Physical Review C {\bf 88},
044618 (2013).
\bibitem{san1}K.P. Santhosh, J.G. Joseph, B. Priyanka, and S. Sahadevan,
J. Phys. G {\bf 38}, 075101 (2011).
\bibitem{lov1}R.G. Lovas, R.J. Liotta, A. Insolia, K. Varga, and D.S.
Delion, Phys. Rep. {\bf 294}, 265 (1998).
\bibitem{car1}N. Carjan, A. Sandulescu, and V. V. Pashkevich,
Phys. Rev. C {\bf 11}, 782 (1975).
\bibitem{mang}H. J. Mang, Phys. Rev. {\bf 119}, 1069 (1960).
\bibitem{san2}A. Sandulescu, Nucl. Phys. {\bf 37}, 332 (1962).
\bibitem{po}D. N. Poenaru and W. Greiner, J. Phys. G {\bf 17}, S443 (1991).
\bibitem{zd}A. Zdeb, M. Warda, and K. Pomorski, Phys. Rev. C {\bf 87},
024308 (2013).
\end{thebibliography}
\end{document}